\documentclass[aps,prd,twocolumn,10pt,showpacs,preprintnumbers,nofootinbib,english,floatfix]{revtex4-1}
\usepackage{times,amsmath,amsfonts,amssymb}
\usepackage[usenames, dvipsnames]{xcolor}
\usepackage[T1]{fontenc}
\usepackage{textcomp}
\usepackage{dcolumn}
\usepackage{bm}
\usepackage{graphicx}
\usepackage{epstopdf, epsfig}
\usepackage{url}
\usepackage[normalem]{ulem}
\usepackage{hyperref}

\hypersetup{colorlinks=true, linkcolor=NavyBlue, citecolor=NavyBlue, bookmarksopen=true,%
			pdfstartview={FitH}, pdfpagelayout=OneColumn, pdfstartpage=1}
\newcommand*{\tstr}{\texorpdfstring}
\newcommand*{\negs}{\negthickspace}
\newcommand*{\td}[1]{\tilde{#1}}
\begin{document}

\title{Can background cosmology hold the key for modified gravity tests?}

\author{Juan~J.~Ceron-Hurtado}

\author{Jian-hua~He}

\author{Baojiu~Li}

\affiliation{Institute for Computational Cosmology, Department of Physics, Durham University, Durham DH1 3LE, UK}

\date{\today}

\hypersetup{bookmarksdepth=-2}
\begin{abstract}

Modified gravity theories are a popular alternative to dark energy as a possible explanation for the observed accelerating cosmic expansion, and their cosmological tests are currently an active research field. Studies in recent years have been increasingly focused on testing these theories in the nonlinear regime, which is computationally demanding. Here we show that, under certain circumstances, a whole class of theories can be ruled out by using background cosmology alone. This is possible because certain classes of models (i) are fundamentally incapable of producing specific background expansion histories, and (ii) said histories are incompatible with local gravity tests. As an example, we demonstrate that a popular class of models, $f(R)$ gravity, would not be viable if observations suggest even a slight deviation of the background expansion history from that of the $\Lambda$CDM paradigm.

\end{abstract}
\hypersetup{bookmarksdepth}


\maketitle

\section{Introduction}

In the past decade or so, alternative theories of gravity as a possible explanation for the accelerating expansion of the Universe have received a great deal of attention \cite{cfps2011,joyce2015}. Such theories affect the dynamics of the expansion on cosmological scales, where General Relativity (GR) is usually assumed to break down, without invoking a mysterious new matter species commonly known as dark energy. Thanks to the development of linear and nonlinear computational tools in recent years, this area has advanced quickly, with the formation of large-scale structures in many of the new models being fairly well understood by now, and the study of baryonic and galaxy evolution in them already initiated by some groups \cite[e.g.,][]{aps2014, hlmw2015, hl2015}.

There are, however, a few challenges hindering further development of the field. Many of the alternative theories, such as $f(R)$ gravity \cite{sf2010,dt2010}, indeed have GR as a limit, which means that there is some point (typically characterised by one or more model parameters) after which the theory is no longer distinguishable from GR in practice. In $f(R)$ gravity, for example, Ref.~\cite{jvj2013} shows that a model parameter, $|f_{R0}|$ (to be explained below), has to be smaller than $\sim10^{-7}$ for it to satisfy astrophysical constraints, thus making the cosmology of the model very similar to the general-relativistic prediction. We therefore face the situation that a cosmological model might never be ruled out by cosmological observations. Adding to this is the fact that studies of nonlinear structure formation in the remaining allowed parameter space are increasingly more challenging with ever higher resolution requirements, and systematics and uncertainties start to dominate over model differences from GR. Hence, it is beneficial to find other, hopefully cleaner, ways of testing the models using cosmology.

One place we can look into, as we shall show below, is background cosmology. This may sound counter-intuitive: after all, given the purpose of modified gravity theories, fitting background cosmology seems to be the first test they need to pass. However, many of these alternative theories are known to have great flexibility -- for example, the fourth-order nature of the $f(R)$ gravity equations means that there is an infinite family of models which can exactly reproduce the background expansion history of the $\Lambda$CDM scenario \cite{shs2007}, thus giving us the freedom to simply adopt this standard background and focus on other effects (e.g., the fifth force) on cosmic structure formation.

In this paper, we revisit the role of background cosmology in constraining modified gravity theories. With $f(R)$ gravity as a working example, we will demonstrate that the model is incapable of reproducing certain expansion histories. Furthermore, we exemplify the restrictions on the expansion history itself brought about by the findings of Ref.~\cite{bbds2008}, namely that for this model to be viable its background cosmology has to be very close to the $\Lambda$CDM prediction. This result is generic and model-independent, as it is not a direct constraint on $f(R)$ model parameters. Therefore, if future observations support a dark energy equation-of-state parameter $w$ that is different from $-1$ or evolves in time, the whole $f(R)$ class of theories as an explanation to the cosmic acceleration could be ruled out. This highlights the importance and potential benefits of employing future background cosmological observations in tests of gravity. 

This paper is organised as follows. In Section II, we give a brief overiew of the theory behind $f(R)$ gravity and the relevant field equations. In Section III, we explain how there are certain expansion histories that cannot be reproduced whatever the functional form of $f(R)$, in spite of the fourth-order nature of the theory. In Section IV, we show how deviations from the $\Lambda$CDM expansion history would require $f(R)$ to take on a form that makes it difficult to satisfy local gravity tests. We then give an example of the constraints that can be placed using these arguments in Section V, and present our conclussions in Section VI.

\section{\tstr{$}{}f(R)\tstr{$}{} Gravity}

$f(R)$ gravity is defined by the modified action
\begin{equation}
S = \int{\rm d}^4x\sqrt{-g}\left[\frac{1}{16\pi G}(R+f(R))+\mathcal{L}_m\right],
\end{equation}
where $G$ is Newton's constant, $g$ the determinant of the metric tensor, $R$ the Ricci scalar and $\mathcal{L}_m$ the matter Lagrangian density. With the addition of the nonlinear function $f(R)$, Einstein's equation becomes
\begin{equation}\label{eq:einstein}
G_{\mu\nu}\!+\!f_RR_{\mu\nu}\!+\!\left[\Box f_R\!-\!\frac{f}{2}\right]\!g_{\mu\nu}\!-\!\nabla_\mu\nabla_\nu f_R = 8\pi GT_{\mu\nu},
\end{equation}
in which $g_{\mu\nu}$, $R_{\mu\nu}$, $G_{\mu\nu}$ and $T_{\mu\nu}$ are respectively the metric, Ricci, Einstein and energy-momentum tensors, $\Box\equiv\nabla^\alpha\nabla_\alpha$, and $f_R\equiv{\rm d}f/{\rm d}R$ is a new dynamical degree of freedom (a scalar field) of this theory. Greek indices $\mu,\nu,\ldots$ run over $0,1,2,3$.

Following \cite{shs2007}, we define the dimensionless variables $E~\equiv~H^2/H_0^2$ and $y\equiv f/H_0^2$, where $H\equiv\dot{a}/a$ is the Hubble expansion rate, $a$ the cosmic scale factor, dot the derivative with respect to cosmic time, and the subscript $_0$ denotes the present-day value of a quantity. Using $E$ and $'\equiv{\rm d}/{\rm dln}a$, the curvature scalar becomes $R=3H_0^2(E'+4E)$. For simplicity, we only consider the matter dominated era and the acceleration phase, so that radiation can be neglected. The modified Friedmann equation can then be cast into the form
\begin{equation}\label{eq:friedmann}
\begin{split}
& y'' - \left[1+\frac{E'}{2E} + \frac{E'''+4E''}{E''+4E'}\right]y' + \frac{E''+4E'}{2E}y\\
& = -\frac{E''+4E'}{E}\frac{8\pi G\rho_{\rm DE}}{H_0^2}.
\end{split}
\end{equation}
In writing the above, we have introduced a dark energy fluid with density $\rho_{\rm DE}(a)$, subject to a flat general-relativistic reference model. In this way, the relationship between $H(a)$ and $\rho_{\rm DE}(a)$ is determined as in GR. We define the dark energy equation-of-state parameter $w(a)$ as usual, so that $\rho_{\rm DE}(a)$ satisfies the standard conservation equation
\begin{equation}\label{eq:econs}
\dot{\rho}_{\rm DE}(a) + 3\big[1+w(a)\big]H(a)\rho_{\rm DE}(a) = 0,
\end{equation}
and the $\Lambda$CDM paradigm features $w(a)=-1$ identically.

It is usually claimed that the fourth-order nature of the derivatives appearing in the modified Einstein equation (\ref{eq:einstein}) endows the theory with the freedom to produce arbitrary background expansion histories associated with dark energy -- all this while remaining consistent with observational bounds and approaching $\Lambda$CDM as a limiting case at both high and low redshifts. Conversely, as done in \cite{shs2007} and the present work, one can solve Eq.~(\ref{eq:friedmann}) to obtain a suitable family of functions capable of reproducing a given expansion history, and choose appropriate initial conditions to pick out a particular functional form of $f(R)$.

\section{Background Expansion History}

In this section, we first check the flexibility of $f(R)$ gravity to produce general expansion histories. For illustration purposes we adopt a specific parameterisation of the dark energy equation-of-state parameter (or equivalently the expansion history) \cite{l2003},
\begin{equation}\label{eq:eos}
w(a) = w_0+w_1(1-a),
\end{equation}
in which $w_0, w_1$ are constants, and $\Lambda$CDM is recovered with $w_0=-1$ and $w_1=0$. Although this formula does not cover all possible expansion histories, we will argue that the result derived using it is generic.  

The usual impression that $f(R)$ gravity can reproduce any expansion history hinges on the ability to tune the functional form of $f(R)$, but the claim that this can {\it always} be done is not necessarily true. For example, if $R$ has an extremum\footnote{We use the convention that $R>0$ today.} $R_{\rm min}$ at some time, e.g. when the scale factor $a=a^\ast$, then $f(R)$ is fully determined by the expansion history at $a<a^\ast$: as soon as $a$ crosses $a^\ast$, $R$ will start retracing the values it took on {\it before} $a^\ast$, and so will $f(R)$ and its derivatives. There is no guarantee that the pre-fixed $f(R)$ can still lead to the desired expansion history at $a>a^\ast$, and there is no freedom left to achieve this through further tuning. In fact, such an extremum will cause Eq.~(\ref{eq:friedmann}) to become singular, since $R'=0$ means $E''+4E'=0$, and this will in turn produce inconsistencies in the evolution of observable quantitites\footnote{One can actually argue that this theoretical difficulty creates a problem for $f(R)$ gravity even if $R'=0$ only in the finite future. But in this paper we shall simply require that $R'=0$ does not happen at $a<1$.} across $a=a^\ast$. Therefore, if $w(a)$ causes $R'$ to cross zero at least once prior to $a=1$, then the expansion history it describes cannot possibly be reproduced by $f(R)$ gravity.

To clarify why this would cause problems in general, we next (i) derive the functional form of $f(R)$ that gives the same expansion history as $w(a)$ {\it before} $a^\ast$, and (ii) use this form to calculate the evolution {\it after} $a^\ast$ and check its consistency.

In order to do (i), we first fix $\rho_{\rm DE}(a)$, $E$ and its derivatives with respect to $N=\ln(a)$. Solving Eq.~(\ref{eq:econs}) and defining
\begin{equation}\label{eq:Dfunc}
D(a)\equiv\exp[-3w_1(1-a)]a^{-3(1+w_0+w_1)},
\end{equation}
allows us to rewrite the standard Friedmann equation as
\begin{equation}\label{eq:DE}
E(a)=\Omega_ma^{-3}+(1-\Omega_m)D(a).
\end{equation}
Since $R(N)$ is monotonic for $N\leq N^\ast=\ln(a^\ast)$, we can then solve Eq.~(\ref{eq:friedmann}) numerically to obtain $y(R)$, or equivalently $f(R)$. Integration is performed from $z_{\rm ini}=150$ -- which is deep in the matter era when radiation and dark energy can be neglected for the purpose of this study -- up to a time slightly earlier than $N^\ast$, where the singularity occurs. To set the initial conditions $y(N_{\rm ini})$ and $y'(N_{\rm ini})$, we note that in this era Eq.~(\ref{eq:friedmann}) can be simplified to
\begin{equation}
y'' + \frac{7}{2}y' - \frac{3}{2}y = 9(1-\Omega_m)D(a).
\end{equation}

The associated homogeneous equation has the solution $y_h~=~A_+y_+ + A_-y_-$, where $A_\pm$ are constant coefficients and $y_\pm = \exp(r_\pm N)$, with $r_\pm = (-7\pm\sqrt{73})/4$. The decaying mode associated with $r_-<0$ is dropped here so as to prevent $f(R)$ from being unbounded at early times. Meanwhile, the particular solution $y_p$ can be written in the form $y_p=u_+y_+ + u_-y_-$, with the derivatives of $u_\pm$ solving the system
\begin{equation}
\left\{\ \begin{aligned}
u_+'y_+ + u_-'y_- &= 0,\\
u_+'y_+' + u_-'y_-' &= 9(1-\Omega_m)D(a),
\end{aligned}\right.
\end{equation}
and leading to
\begin{equation}
u_\pm(a) = \pm\frac{9(1-\Omega_m)}{r_+-r_-}\int^a\negs\frac{D(\td{a})}{y_\pm(\td{a})}\frac{{\rm d}\td{a}}{\td{a}}.
\end{equation}
The initial conditions at $a_{\rm ini}=1/(1+z_{\rm ini})$ are derived as
\begin{equation}
\begin{aligned}
y(N_{\rm ini}) &= Ay_+(N_{\rm ini}) + u_-(N_{\rm ini})\,y_-(N_{\rm ini}),\\
y'(N_{\rm ini}) &= Ay'_+(N_{\rm ini}) + u_-(N_{\rm ini})\,y'_-(N_{\rm ini}),
\end{aligned}
\end{equation}
where we have defined $A\equiv A_+ + u_+(N_{\rm ini})$, and it should be noted that the decaying mode $y_-$ has reemerged through the form of the particular solution.

\begin{figure}
\includegraphics[scale=0.45]{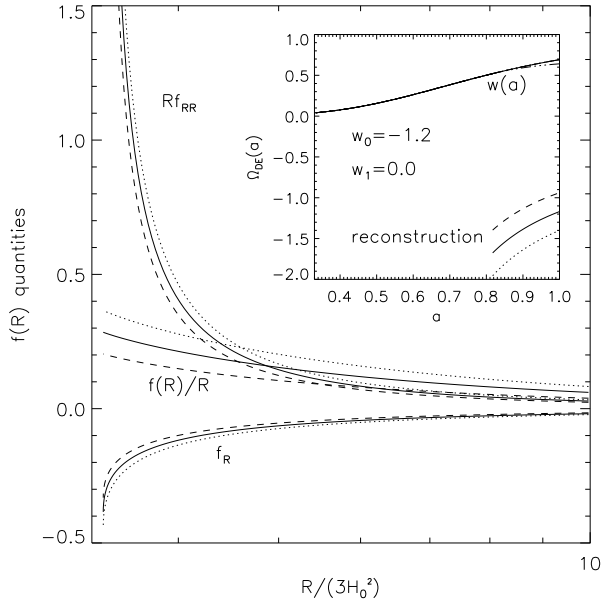}
\caption{\label{fig:fders}{\it Main panel}: Dimensionless quantities $f(R)/R$, $f_R$, $Rf_{RR}$ as functions of $R/3H_0^2$, derived by requiring the expansion history to match that of dark energy with $w(a)=-1.2$ for $a<a^\ast\approx0.817$. The solid, dotted and dashed curves correspond to $A=0,+1$ and $-1$ respectively (c.f. main text for definition of $A$). The result depends on $A$, but the essential feature that $f_{RR}$ diverges at $R_{\rm min}=R(a^\ast)$ is generic.
{\it Insert}: $\Omega_{\rm DE}(a)$ for dark energy with $w(a)=-1.2$ (top solid line) and from the reconstructed $f(R)$ (bottom right; solid, dotted and dashed lines for $A=0,+1$ and $-1$ respectively). Note that $f(R)$ is fully determined by the expansion history for $a<a^\ast$, when both approaches give the same result. However, the pre-fixed $f(R)$ gives rise to a reconstructed $\Omega_{\rm DE}(a)$ {\it after} $a^\ast$ that is incompatible with the curve deduced from $w(a)$, even though one can make it continuous at $a^\ast$ by tuning $A$ (broken line at top right, which corresponds to $A\approx-7.858$).}
\end{figure}

As a concrete example, consider the case with $w_0=-1.2$ and $w_1=0$, and let the present matter fractional density be $\Omega_m=0.308$ \cite{planck2015}. It can be checked that $R'\propto E''+4E'$ vanishes at a time $N^\ast\approx-0.202<0$. As shown in the main panel of Fig. \ref{fig:fders}, integrating Eq.~(\ref{eq:friedmann}) yields a diverging $f_{RR}$ due to the singularity at $N^\ast$, even though $R, f, f_R$ are all continuous at $R_{\rm min}=R(N^\ast)$.

Now, in order to do (ii), we cast Eq.~(\ref{eq:friedmann}) into a more physical form,
\begin{equation}\label{eq:friedmann2}
f_R(H^2+HH') - \frac{1}{6}f(R) - H^2f_{RR}R' = \frac{8\pi G}{3}\rho_{\rm DE}(a).
\end{equation}
As $a$ grows beyond $a^*$, $R$ consecutively revisits the values it took on at $a<a^\ast$, allowing us to substitute the already-fixed forms of $f, f_R, f_{RR}$ into Eq.~(\ref{eq:friedmann2}) and reconstruct\footnote{As a consistency check, $H, H', R'$ assume the forms determined by $w(a)$.} $\rho_{\rm DE}(a)$ for $a>a^*$. The resulting $\Omega_{\rm DE}\equiv8\pi G\rho_{\rm DE}/3H^2$ as a function of time is depicted in the insert of Fig.~\ref{fig:fders}, wherein we clearly spot an unphysical discontinuity arising from the fact that the behaviour of $\Omega_\text{DE}(a)$ at $a<a^*$, which was obtained from the solution to Eq.~(\ref{eq:friedmann}) with equation of state (\ref{eq:eos}), is incompatible with the reconstruction at $a>a^*$ built from the previously fixed form of $f(R)$.

This conclusion holds for general forms of $w(a)$: if the desired expansion history has $R'$ crossing zero before $a=1$, it {\it cannot} be produced by $f(R)$ gravity.  Later, we will constrain $(w_0,w_1)$ by requiring that $R'=0$ never happens at $a\leq1$.

\section{Chameleon Screening}

Chameleon screening \cite{kw2004} has been an active research topic recently, and it is what enables $f(R)$ gravity to potentially evade stringent solar system tests \cite{will2014}. This can be seen from the modified Einstein equation (\ref{eq:einstein}), which shows that the theory goes back to GR in the limit that $f_R(R)\rightarrow0$ for a wide range of $R$ values. This means that $f'=f_RR'$ will be small, and similarly will $f''$, so that from Eq.~(\ref{eq:friedmann}) one has $f\propto\rho_{\rm DE}(a)$ $\sim{\rm const.}$ approximately. Therefore, for the theory to employ the chameleon mechanism to evade local constraints, $f$ should depend weakly on $R$ ($|f_R|\ll1$), and so the expansion history must be close to that of $\Lambda$CDM\footnote{One can also see this directly from Eq.~(\ref{eq:einstein}), in which the term $fg_{\mu\nu}$ is what drives the accelerated expansion. Between $z=0$ and $1$, $R$ changes by $\sim$10 for a typical expansion history. For $|f_R|<\mathcal{O}(10^{-5})$ required to pass solar system gravity tests, $f$ changes by $\Delta f/f \sim f_R\Delta R/f \sim10^{-4}$ (where we assume $|f|\sim|R_0|$) and so acts practically as a cosmological constant.}. The idea has previously been described in \cite{bbds2008}, where the authors also demonstrate it using several explicit forms of $f(R)$ (see \cite{vds16} for an example where, using full numerical simulations, the screening is shown not to work in $f(R)$ gravity when $w(a)$ deviates substantially from $-1$). Here we illustrate this from a different angle -- by finding stringent restrictions on the $w_0$-$w_1$ parameterisation based on the screening requirement.

\begin{figure*}
\includegraphics[scale=0.45]{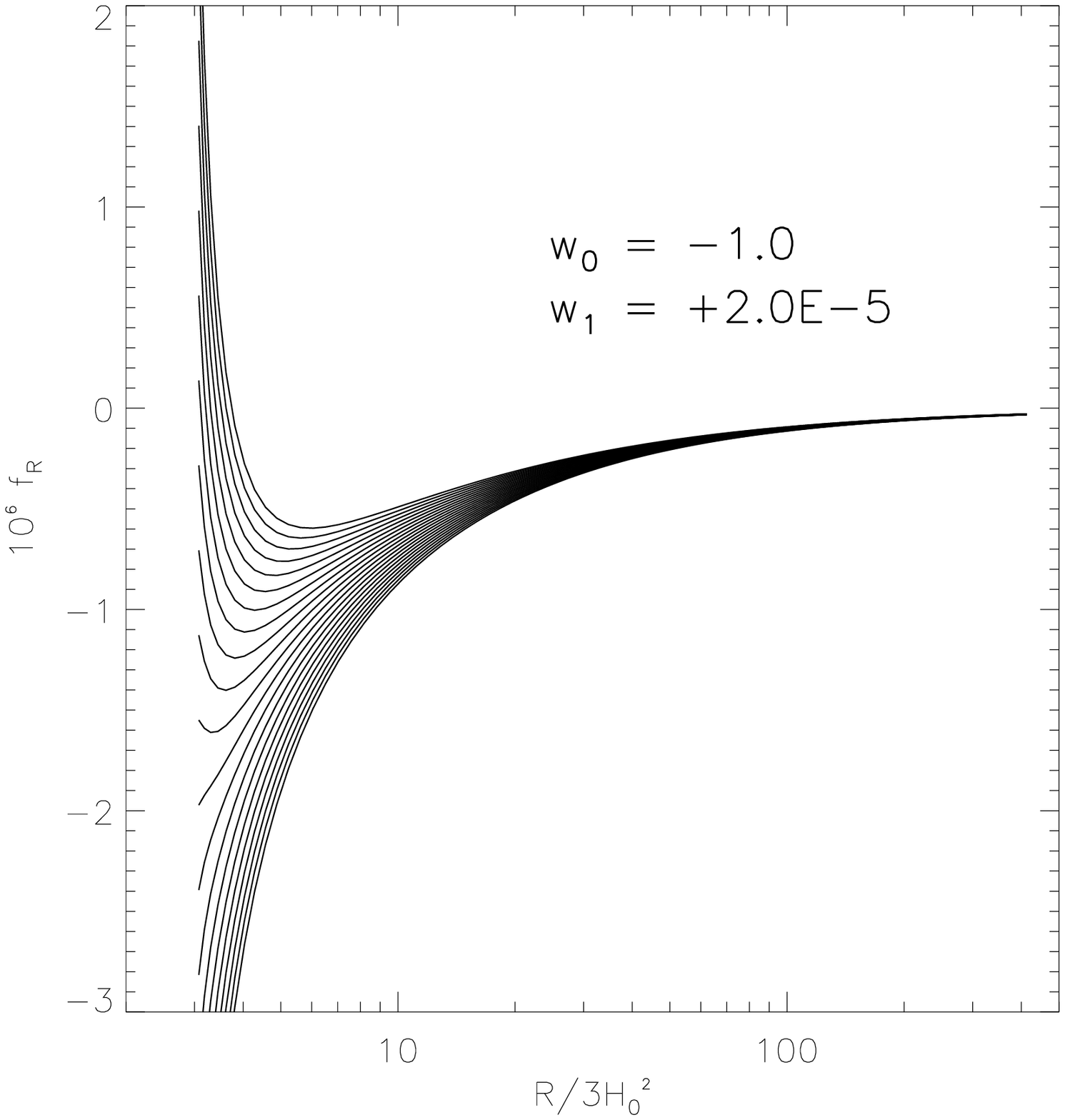}
\includegraphics[scale=0.45]{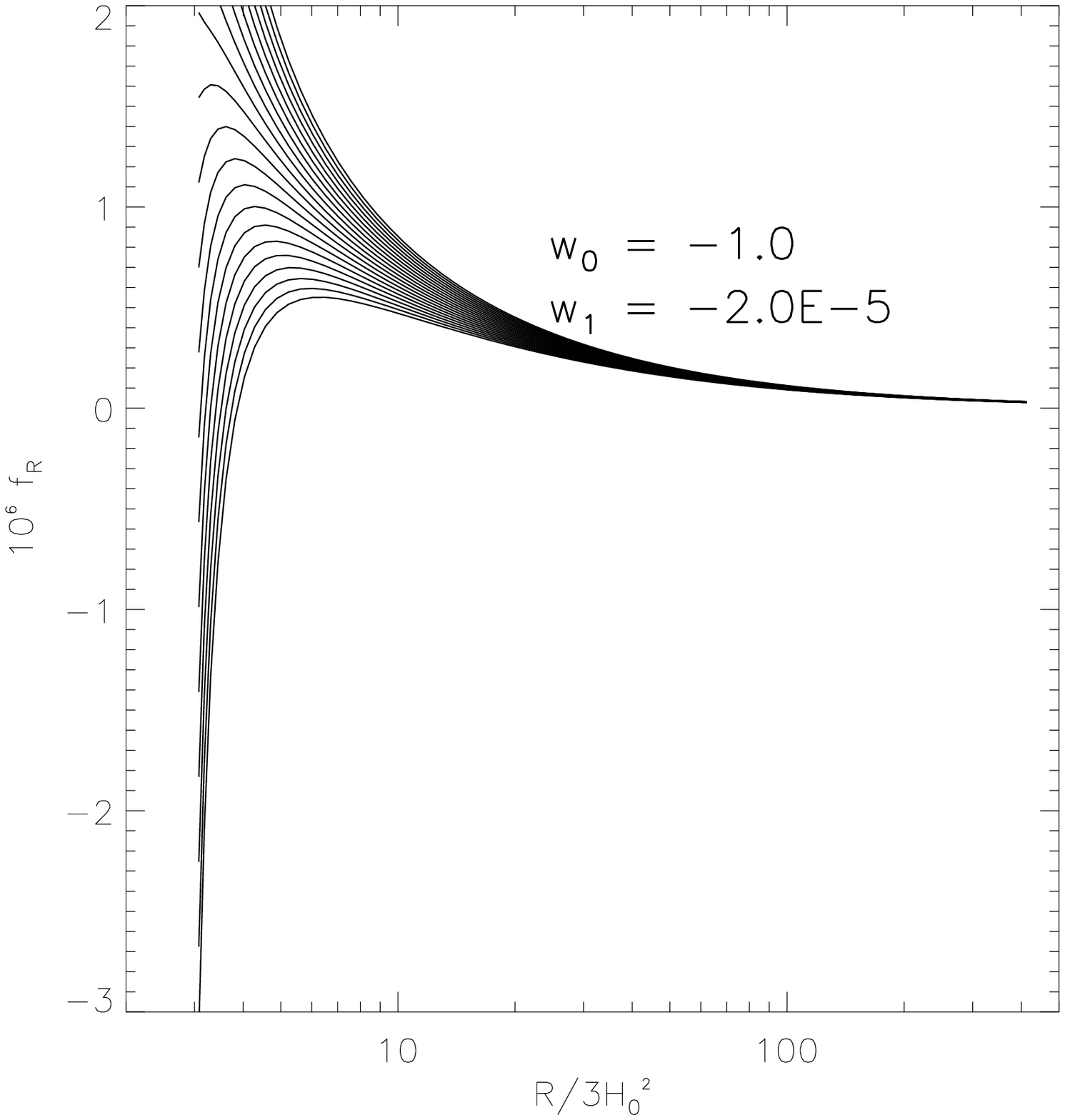}
\caption{\label{fig:fR_evolution}Examples of the reconstructed $f_R$ as a function of $R$ for a background expansion history matching that of dark energy with $w_0=-1$ and $|w_1|=2\times10^{-5}$. The evolution with a positive value of $w_1$ is shown on the left panel, while that with negative $w_1$ appears on the right. The different curves correspond to different values of $A$ (or equivalently $f_{R0}$) and $\Omega_m=0.308$ in all cases. Depending on $A$, $f_R$ may display an extremum, vary by more than $2\times 10^{-6}$ in magnitude over the whole curvature range, or both.}
\end{figure*}

To see why the solar system constraints imply a small $|f_R|$, let us note that the efficiency of chameleon screening can be neatly characterised by the so-called thin-shell condition \cite{kw2004},
\begin{equation}
\frac{\Delta r}{r} \sim \frac{|f_{R,\rm in}-f_{R,\rm out}|}{\Phi_N},
\end{equation}
for a spherical body with a top-hat density profile. Here, $r$ is the radius of the body, $\Phi_N$ the Newtonian potential it creates at its surface, and $f_{R,{\rm in}}$, $f_{R,{\rm out}}$ respectively the values of the scalar field, $f_R$, that minimise the effective potential inside and outside the body. $\Delta r \leq r$ is the thickness of a shell, the matter within which produces a fifth force that has $1/3$ of the strength of Newtonian gravity. To be compatible with solar system gravity tests, the fifth force must be weak (decay rapidly with distance), which requires $\Delta r \ll r$. Generally, $|f_{R,\rm in}|\ll|f_{R,\rm out}|$ because $|f_R|$ decreases with growing $R$, and so the above condition translates into $|f_{R,\rm out}|\ll|\Phi_N|$ with $|\Phi_N|\lesssim10^{-4}$ for cosmological and astrophysical bodies. The strongest constraint to date comes from astrophysical considerations and sets $|f_{R0}|\lesssim10^{-7}$ \cite{jvj2013}, while more modest limits of $10^{-6}\sim10^{-4}$ are deduced mainly from cosmology by various groups (e.g., \cite{Cataneo2015,Liu2016}).

Figure \ref{fig:fR_evolution} shows $f_R(R)$ for a background expansion history that matches that of a dark energy model with $w_0=-1$ and $|w_1|=2\times10^{-5}$, which is very close to $\Lambda$CDM; the left panel corresponds to $w_1>0$, and the right panel to $w_1<0$. The result depends qualitatively on $A$ in both situations, but there are two noticeable features they have in common:

Firstly, $f_R(R)$ has an extremum for certain values of $A$, namely a minimum if $w_1>0$ or a maximum if $w_1<0$. This means that it is not possible to invert it to find $R(f_R)$ in either case. Since $f_R$ plays the role of a scalar field whose potential satisfies \cite{lb2007,hs2007,bbds2008}
\begin{equation}
\frac{\partial V(f_R)}{\partial f_R} = \frac{1}{3}\big[(1-f_R)R(f_R)+2f(f_R)-8\pi G\rho_m\big],
\end{equation}
it follows that the potential cannot be uniquely defined. Whilst this does not necessarily mean that the theory itself is ill-defined (a question we do not attempt to tackle here), it does make it difficult to envisage how the chameleon mechanism could work to suppress deviations from GR inside the solar system.

Secondly, and more importantly, we see that even for values of $A$ such that $f_R(R)$ is monotonic -- increasing if $w_1$ is positive but decreasing if it is negative -- the field magnitude $|f_R|$ generally changes by more than $2\times10^{-6}$ from high curvature ($R\gg3H_0^2$) to low curvature ($R\sim3H_0^2$) in this specific case. This strong variation of $f_R$ with $R$ prevents the chameleon mechanism from being efficient enough to pass solar system tests if we take the constraint $|f_{R0}|=|f_R(R_0)|\lesssim10^{-6}$ at face value. Indeed, from Eq.~(\ref{eq:einstein}) we see that non-GR terms such as $\nabla^2f_R$ can be of the same order as, or even larger than, the standard Newtonian term $\nabla^2\Phi$ for $|\Phi|\lesssim10^{-6}$. An additional concern is that a decreasing $f_R$, either monotonically or in certain ranges of $R$ (and time), would make $f_{RR}<0$ and lead to unstable growth of $f_R$ perturbations \cite{shs2007,lb2007}.

As mentioned above, although our discussion is based on a specific non-$\Lambda$CDM expansion history governed by the equation of state~(\ref{eq:eos}), we can claim in general that in order for the chameleon screening to be efficient in other expansion histories, $f_R$ cannot change significantly from high- to low-curvature regions, thus requiring $f(R)$ to remain nearly constant at low redshift. One can easily put certain classes of $f(R)$ models into this context; the most popular example is that of \cite{hs2007}, for which $f_R$ is monotonic in $R$. It is known that in this model ($n=1$) one needs $|f_{R0}|\lesssim10^{-6}$ to pass solar system tests, which in turn hinges on the expansion history being extremely close to that of $\Lambda$CDM (cf.~Fig.~3 there). We note that other studies have also pointed out that viable $f(R)$ models must behave very similarly to $\Lambda$CDM. In particular, Ref. \cite{cd2015} finds that the model would otherwise admit no observationally viable weak field limit.

\section{Results}

We shall now constrain the parameter space $(w_0,w_1)$ by requiring that $R'$ not cross zero at $a\leq1$ (cf.~Section III), that $f_R$ increase monotonically with $R$, and that $|f_R|$ not vary by over $10^{-6}$ in the whole $R$ range (cf.~Section IV). This is not a full calculation (which would involve numerical simulations, e.g. \cite{o2008,lz2009,zlk2011,ecosmog}, to properly model the environmental effects on the screening), but rather a rough estimation to show how close to $\Lambda$CDM the expansion history needs to be.

\begin{figure}
\includegraphics[scale=0.45]{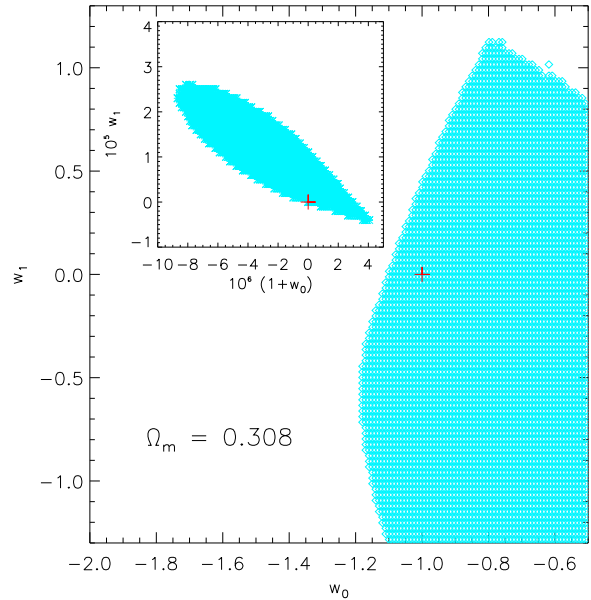}
\caption{\label{fig:countour}(Color online) {\it Main panel}: Region of the $w_0$-$w_1$ plane where $R(a)$ has no extrema at $a\leq1$, allowing the specified expansion history to be consistently reproduced by $f(R)$ gravity. {\it Insert}: Region where the reconstructed $f(R)$ model can pass solar system tests (see the main text for more details). Red crosses correspond to $\Lambda$CDM.}
\end{figure}

The main panel of Fig.~\ref{fig:countour} shows the region of the $w_0$-$w_1$ parameter space where $R(a)$ has no extrema at $a\leq1$ and it is possible to use $f(R)$ gravity to reproduce the whole expansion history (up to $a=1$). As expected, $\Lambda$CDM, with $(w_0,w_1)=(-1,0)$, can be exactly mimicked by this framework \cite{hw2013}. However, part of the parameter space allowed by CMB data and other probes \citep[][Fig.~28]{planck2015} does not correspond to viable $f(R)$ models.

A much stronger constraint comes from the requirement of passing solar system tests \cite{bbds2008}, as shown in the insert panel of Fig.~\ref{fig:countour}. We find that neither $w_0$ nor $w_1$ can deviate by more than $\mathcal{O}(10^{-5})$ in magnitude from its $\Lambda$CDM value, if the perturbation evolution of $f_R$ is to be stable and the chameleon screening is to remain efficient.

\section{Discussion and Conclusions}

The results indicate a special property of $f(R)$ gravity, namely that, for it to be viable, the expansion history cannot be arbitrary but has to be very close to $\Lambda$CDM. More precisely, parameter values $(w_0,w_1)$ in the dark energy equation-of-state $w(a)$ shown in Eq.~(\ref{eq:eos}) {\it cannot} differ from $(-1,0)$ by more than $\mathcal{O}(10^{-5})$, if $f(R)$ gravity is the underlying gravity model. Likewise, any observational evidence for a significant time evolution of $w(a)$ would rule out $\Lambda$CDM and the entire $f(R)$ class of models simultaneously.

For this reason, it is crucial to further improve the observational constraints on $w(a)$. Most current limits do not rule out $w(a)=-1$, though in some cases it is not the best fit \cite{zcpz2012}. Future galaxy surveys, such as Euclid \cite{euclid} and {\sc desi} \cite{desi}, have the potential of reducing the uncertainty on $w_0$ and $w_1$ to $\Delta w_0\sim0.01$ and $\Delta w_1\sim0.05$. This means that small deviations -- if they exist -- from $w=-1$ can be measured, and in turn be used to rule out $f(R)$ gravity. One caveat, however, is that these estimates often stem from the synergy of different probes or even different surveys (e.g., including Planck): if the probes rely on the growth rate of matter perturbations, the derived constraints on $w_0, w_1$ depend on the gravity model (which in most forecasts is taken to be GR), and can't be used universally. Geometric measures, such as the baryon acoustic oscillation peak positions, could be used in a more model-independent way, but on their own the constraints would be weaker.

Although most research efforts so far have focused on the goodness-of-fit of $w(a)$ parameterisations, the dark energy equation of state can indeed also be reconstructed nonparametrically with the fewest possible assumptions. Ref.~\cite{zcpz2012} contains such an example of reconstruction, where it is found that a time-varying $w(a)$ is slightly preferred over the $\Lambda$CDM case. Such work can prove invaluable in constraining theories like $f(R)$ gravity.

From a more general point of view, $f(R)$ gravity is a subclass of the chameleon theory \cite{kw2004}, so its screening mechanism relies similarly on the scalar field staying small from very high to low curvatures, or equivalently from very early times to today \cite{bdl2012}. This means that the scalar field will barely evolve and its potential energy, which drives the cosmic acceleration, will stay nearly constant in time: once more, we have a background expansion history that has to be close to $\Lambda$CDM, as generically described by \cite{bbds2008}.

To summarise, alternative theories of gravity have been extensively studied in the past few years due to their potential to explain the accelerated cosmic expansion. Although such theories have rich phenomenology in terms of structure formation, and can therefore be constrained using observations associated with the latter, we argue that in certain cases the expansion history itself may be used as a smoking gun to rule out classes of theories. This is based on two observations: first, a theory could be intrinsically incapable of producing certain background expansion histories, by analogy with the observation that quintessence models cannot produce a phantom ($w<-1$) background evolution; second, a theory capable of producing a certain expansion history might yield unexpected and unwanted phenomena at small scales. We show that both possibilities indeed happen in one of the most popular theories -- $f(R)$ gravity or the chameleon theory -- leading to the strong constraint that the expansion history must be very close to that of $\Lambda$CDM. Consequently, precise measurements of the dark energy equation of state, $w(a)$, can be useful in ruling out such theories in the future. In contrast, studies of structure formation, which have been the main focus in recent times, are more likely to result in reduced parameter spaces rather than the exclusion of whole classes of models.


\begin{thebibliography}
\bibitem{} \ifx\csname natexlab\endcsname\relax \fi \expandafter\ifx\csname
bibnamefont\endcsname\relax

\fi \expandafter\ifx\csname bibfnamefont\endcsname\relax

\fi \expandafter\ifx\csname citenamefont\endcsname\relax

\fi \expandafter\ifx\csname url\endcsname\relax

\fi \expandafter\ifx\csname urlprefix\endcsname\relax

\fi \providecommand{\bibinfo}[2]{#2} \providecommand{\eprint}[2][]{\url{#2}}


\bibitem{cfps2011}  T.~Clifton, P.~G.~Ferreira, A.~Padilla and C.~Skordis, Phys.~Rept.,  {\bf 513}, 1 (2012).

\bibitem{joyce2015}  A.~Joyce, B.~Jain, J.~Khoury and M.~Trodden, Phys.~Rept., {\bf 568}, 1 (2015).

\bibitem{aps2014} C.~Arnold, E.~Puchwein and V.~Springel, Mon.~Not.~R.~Astron.~Soc., {\bf 440}, 833 (2014).

\bibitem{hlmw2015} A.~Hammami, C.~Llinares, D.~F.~Mota and H.~A.~Winther, Mon.~Not.~R.~Astron.~Soc., {\bf 449}, 3635 (2015).

\bibitem{hl2015} J.~He and B.~Li, Phys.~Rev.~D{\bf93}, 123512 (2016).

\bibitem{sf2010} T.~P.~Sotiriou and V.~Faraoni, Rev.~Mod.~Phys., {\bf82}, 451 (2010).

\bibitem{dt2010} A.~De~Felice and S.~Tsujikawa, Living Rev.~Rel., {\bf 13}, 3 (2010).

\bibitem{jvj2013} B.~Jain, V.~Vikram and J.~Sakstein, Astrophys.~J. {\bf 779}, 39 (2013).

\bibitem{shs2007} Y.~Song, W.~Hu and I.~Sawicki, Phys.~Rev.~D{\bf75}, 044004 (2007).

\bibitem{bbds2008} P.~Brax, C.~van~de~Bruck, A.~C.~Davis and D.~J.~Shaw, Phys.~Rev.~D{\bf78}, 104021 (2008).

\bibitem{l2003} E.~V.~Linder, Phys. Rev. Lett.~{\bf 90}, 091301 (2003).

\bibitem{planck2015} Planck Collaboration: P.~A.~R.~Ade {\it et al.}, arXiv:1502.01589.

\bibitem{kw2004} J.~Khoury and A.~Weltman, Phys.~Rev.~D{\bf69}, 044026 (2004).

\bibitem{will2014} C.~M.~Will, Living Rev.~Rel., {\bf 17}, 4 (2014).

\bibitem{vds16} M.~Vargas dos Santos, H.~A.~Winther, D.~F.~Mota and I.~Waga, A~\&~A, {\bf 587}, 132 (2016).

\bibitem{Cataneo2015} M.~Cataneo {\it et al.}, Phys.~Rev.~D{\bf 92}, 044009 (2016).

\bibitem{Liu2016} X.~Liu {\it et al.}, Phys.~Rev.~Lett., in press; arXiv:1607.00184 [astro-ph.CO].

\bibitem{lb2007} B.~Li and J.~D.~Barrow, Phys.~Rev.~D{\bf 75}, 084010 (2007).

\bibitem{hs2007} W.~Hu and I.~Sawicki, Phys.~Rev.~D{\bf 76}, 064004 (2007).

\bibitem{cd2015} T.~Clifton and P.~K.~S.~Dunsby, Phys.~Rev.~D{\bf 91}, 103528 (2015).

\bibitem{o2008} H.~Oyaizu, Phys.~Rev.~D{\bf78}, 123523 (2008).

\bibitem{lz2009} B.~Li and H.~Zhao, Phys.~Rev.~D{\bf80}, 044027 (2009).

\bibitem{zlk2011} G.~Zhao, B.~Li and K.~Koyama, Phys.~Rev.~D{\bf83}, 044007 (2011).

\bibitem{ecosmog} B.~Li, G.~Zhao, R.~Teyssier and K.~Koyama, J.~Cosmo.~Astropart.~Phys., {\bf01}, 051 (2012).

\bibitem{hw2013} J.~He and B.~Wang, Phys.~Rev.~D{\bf 87}, 023508 (2013). 

\bibitem{euclid} R.~Laureijs~{\it et al.}, arXiv:1110.3193 [astro-ph.CO].

\bibitem{desi} M.~Levi~{\it et al.}, arXiv:1308.0847 [astro-ph.CO].

\bibitem{zcpz2012} G.~Zhao, R.~G.~Crittenden, L.~Pogosian and X.~Zhang, Phys. Rev.~Lett., {\bf 109}, 171301 (2012).

\bibitem{bdl2012} P.~Brax, A.~C.~Davis and B.~Li, Phys.~Lett.~{\bf B715}, 38 (2012).


\end{thebibliography}
\end{document}